\documentclass[prd,twocolumn,footinbib]{revtex4}

\usepackage{amssymb}
\usepackage{graphicx}
\newcommand{\be}{\begin{equation}}
\newcommand{\ee}{\end{equation}}
\newcommand{\bea}{\begin{eqnarray}}
\newcommand{\eea}{\end{eqnarray}}

\newcommand{\kg}{{k_\mathrm{g}}}
\newcommand{\kd}{{k_\mathrm{d}}}

\begin{document}

\title{
Observational Constraints on the Averaged Universe
  }

\author{
Chris~Clarkson$^{1}$,
Timothy~Clifton$^{2}$,
Alan~Coley$^{3}$
and Rockhee~Sung$^{1}$
\\
\vspace{5pt}
$^{1}${\it {\small Astrophysics, Cosmology \& Gravity Centre, and,
    Department of Mathematics \& Applied Mathematics, University of
    Cape Town, South Africa.}}\\
$^{2}${\it {\small Department of Astrophysics, University of Oxford, UK.}}\\
$^{3}${\it {\small Department of Mathematics and Statistics, Dalhousie
  University, Halifax, Canada.}}\\
}


\begin{abstract}

Averaging in general relativity is a complicated operation, due to the
general covariance of the theory and the non-linearity of Einstein's
equations.  The latter of these ensures that smoothing spacetime over
cosmological scales does not yield the same result as solving
Einstein's equations with a smooth matter distribution, and that the
smooth models we fit to observations need not be simply related to the
actual geometry of spacetime.  One specific consequence of this is a
decoupling of the geometrical spatial  curvature term in the metric from the
dynamical spatial curvature in the Friedmann equation. Here we investigate the
consequences of this decoupling by fitting to a combination of HST,
CMB, SNIa and BAO data sets.   We find that only the geometrical
spatial curvature is tightly constrained, and that our ability to
constrain dark energy dynamics will be severely impaired until we gain
a thorough understanding of the averaging problem in cosmology.

\end{abstract}

\maketitle

\section{Introduction}

The standard model of cosmology relies on the assumption that the
Universe is well described at all points in space and time by a single linearly
perturbed Friedmann-Lema\^{i}tre-Robertson-Walker (FLRW) geometry (except in the vicinity of black holes and
neutron stars), and that this geometry obeys Einstein's equations.
However, in an inhomogeneous universe this is almost certainly not
true.  Even if the gravitational interaction behaves exactly as
Einstein predicted on small scales, this will not be true for
large-scale averages.  A critical problem in cosmology is therefore determining
the form of deviations from Einstein's equations when considering
geometry averaged on large scales, and what the effects of these deviations will be
on observations (for a review, see \cite{Clarkson:2011zq}).  This has been the subject of some controversy,
with opinions ranging from the suggestion that the effects of
averaging could completely explain
the recently observed accelerating expansion of the Universe without
the need for any dark energy
\cite{Buchert:1999er,Buchert:2001sa,Rasanen:2003fy,Barausse:2005nf,Kolb:2005da,Rasanen:2006kp,Kasai:2007fn,Larena:2008be},
to the claim that it is completely negligible
\cite{Ishibashi:2005sj,Flanagan:2005dk,Hirata:2005ei,Geshnizjani:2005ce,Bonvin:2005ps,Bonvin:2006en,
Behrend:2007mf,Kumar:2008uk,Krasinski:2009qq,
Tomita:2009ar,Baumann:2010tm,Green:2010qy}.  Others suggest that
while the effects of averaging may not be responsible for the apparent
acceleration, they may be important for precision cosmology 
\cite{Russ:1996km,Kolb:2004am,Coley:2005ei,Li:2007ci,COLEYAV,Li:2007ny,Li:2008yj,Clarkson:2009hr,
Clarkson:2009jq,Chung:2010xx,Umeh:2010pr}.

In this paper we take the solutions to the field equations
derived using an exact and fully covariant averaging
procedure, and compare them to
observations.   These solutions have decoupled spatial curvature
parameters in the metric and the Friedmann equation, and reduce to the
FLRW solutions of Einstein's equations when these parameters are
equal.   We find that that the
constraints available on the spatial curvature parameter appearing in
the Friedmann equation are considerably weaker than those available on the
spatial curvature parameter appearing of the macroscopic metric. In
particular, the constraints from the Cosmic Microwave Background (CMB) are considerably weakened, and
no longer signal a flat universe.  This allows for the 
possibility of averaging having non-negligible dynamical
consequences.  We also find that some data sets prefer models in which
the two curvature parameters are not equal.

\section{Spacetime Averaging and Macroscopic Gravity}

There are a number of averaging procedures that have been introduced
in order to study the large-scale evolution of inhomogeneous
spacetimes.  An exact and covariant approach that allows tensor
quantities to be averaged, as well as scalars, was provided by
Zalaletdinov \cite{Zala}.  Here the geometric objects that exist on the
spacetime manifold are averaged, and the field equations that these
quantities satisfy are constructed.  This is achieved
using bilocal averaging operators over closed regions of
spacetime, $\Sigma$, that contain the supporting points $x$
(see \cite{Zala}).  The result of averaging $X$ is then denoted by~$\left< X \right>$.

Using this definition we can now consider the average of various
geometric objects.  
Following Zalaletdinov, we denote the average of the connection as
$\left< \Gamma^{\mu}_{\phantom{\mu} \nu \rho} \right>$, and define a
new macroscopic Riemann tensor
\bea
&&M^{\mu}_{\phantom{\mu} \nu \alpha \beta} = \partial_{\alpha}
\left<\Gamma^{\mu}_{\phantom{\mu} \nu \beta}\right>- \partial_{\beta}
\left<\Gamma^{\mu}_{\phantom{\mu} \nu \alpha}\right> 
\nonumber\\&&~~~~~~~~~~~~~
+ \left<\Gamma^{\mu}_{\phantom{\mu}
  \sigma \alpha}\right> \left<\Gamma^{\sigma}_{\phantom{\sigma} \nu \beta}\right> -
\left<\Gamma^{\mu}_{\phantom{\mu} 
  \sigma \beta}\right> \left<\Gamma^{\sigma}_{\phantom{\sigma} \nu
  \alpha}\right>.
\eea
Crucially, $M^{\mu}_{\phantom{\mu} \nu \alpha \beta} \neq \left<
R^{\mu}_{\phantom{\mu} \nu \alpha \beta} \right>$, where $\left<
R^{\mu}_{\phantom{\mu} \nu \rho \sigma} \right>$ is the average of the
microscopic Riemann tensor.  From these quantities one can then
construct the macroscopic field equations \cite{Zala}:
\bea
\label{MFE}
&&\left<g^{\beta \epsilon } \right>M_{\gamma \beta} - \textstyle\frac{1}{2}
\delta^{\epsilon}_{\phantom{\epsilon} \gamma} \left< g^{\mu \nu}
\right> M_{\mu \nu} \\&&~~~~~~
= 8 \pi G \left< T^{\epsilon}_{\phantom{\epsilon}
  \gamma} \right> + \left<g^{\mu \nu} \right>\left(Z^{\epsilon}_{\phantom{\epsilon} \mu \nu
  \gamma} - \textstyle\frac{1}{2} \delta^{\epsilon}_{\phantom{\epsilon} \gamma}
Z^{\alpha}_{\phantom{\alpha} \mu \nu \alpha} \right) ,\nonumber
\eea
where $\left< T^{\epsilon}_{\phantom{\epsilon} \gamma} \right>$ is the
averaged energy-momentum tensor, $\left<g_{\mu \nu} \right>$ is the
averaged metric, and $Z^{\alpha}_{\phantom{\alpha}
  \mu \nu \beta}=2 Z^{\alpha \phantom{\mu [ \epsilon}
    \epsilon}_{\phantom{\alpha} \mu [
      \epsilon \phantom{\epsilon} \underline{\nu} \beta ]}$ is a
  correlation tensor defined by
\be
Z^{\alpha \phantom{\beta [ \gamma} \mu}_{\phantom{\alpha} \beta[
      \gamma \phantom{\mu} \underline{\nu} \sigma ]}
=
\left< \Gamma^{\alpha}_{\phantom{\alpha} \beta [ \gamma}
  \Gamma^{\mu}_{\phantom{\mu} \underline{\nu} \sigma ]} \right> -
\left< \Gamma^{\alpha}_{\phantom{\alpha} \beta [ \gamma} \right>
  \left<\Gamma^{\mu}_{\phantom{\mu} \underline{\nu} \sigma ]} \right>,
\ee
where underlined indices are not included in anti-symmetrization.
This quantity must obey the differential constraint $Z^{\alpha
  \phantom{\beta [ \gamma} \mu}_{\phantom{\alpha} \beta[ \gamma
      \phantom{\mu} \underline{\nu} \sigma ;\lambda]}=0$,
where the covariant derivative here is defined using the averaged
connection $\left< \Gamma^{\mu}_{\phantom{\mu} \nu \rho} \right>$.
The macroscopic field equations (\ref{MFE}) replace Einstein's
equations on large scales.

\subsection{Macroscopic FLRW Solutions}

The solutions to the macroscopic field equations (\ref{MFE}), with an
FLRW ansatz for the macroscopic metric, have recently been studied in
\cite{Coley:2005ei,Coley:2006xu,Coley:2006kp,vandenHoogen:2009nh,vandenHoogen:2009zz},
where it was found that the extra terms involving the correlation tensor $Z^{\alpha \phantom{\beta [ \gamma}
    \mu}_{\phantom{\alpha} \beta[\gamma \phantom{\mu} \underline{\nu}
      \sigma ]}$ take the same form in the macroscopic field equations
  that a spatial curvature 
curvature term takes in Einstein's equations.  In fact,
for a spatially flat macroscopic metric, and with spatial
correlations only, it can be shown that extra terms in Eq. (\ref{MFE})
can {\it only} take the form of spatial curvature.  Any other form
would be incompatible with either the conservation equations,
their integrability conditions, or the algebraic constraints that
$Z^{\alpha \phantom{\beta [ \gamma} \mu}_{\phantom{\alpha} \beta[
      \gamma \phantom{\mu} \underline{\nu} \sigma ]}$ must satisfy for
consistency of the averaging scheme.  Thus, \emph{the averaged
  Einstein field equations for 
a spatially flat, homogeneous, and isotropic macroscopic spacetime
geometry take the form of the Friedmann equations of general
relativity for a non-flat FLRW geometry.} 
That is, the spatial curvature of the macroscopic spacetime
is decoupled from the spatial curvature that appears in the
macroscopic Friedmann equation.  This is an important difference from
the standard 
approach to cosmology, where it is assumed that Einstein's field
equations hold whatever the smoothing scale, and that the spatial curvature in
the Friedmann equation is therefore identical to the spatial curvature
of the macroscopic spacetime.

Using the results above we motivate the following phenomenological cosmological model.
We write the line-element of the macroscopic geometry as
\be
\label{MLE}
ds^2 = \left< g_{\mu \nu} \right> dx^{\mu} dx^{\nu} = -dt^2 + a^2(t) \left[
  \frac{dr^2}{1-\kg r^2} + r^2 d\Omega
  \right],
\ee
where the geometrical curvature, $\kg$, is, in general, a function of
the scale of $\Sigma$.  The scale factor $a(t)$ is that of the
macroscopic spacetime.  On scales larger than $\Sigma$ the macroscopic
field equations, (\ref{MFE}), then become 
\be
\label{FE}
H^2=\frac{\dot{a}^2}{a^2} = \frac{8 \pi G}{3} \rho -\frac{\kd}{a^2}+
\frac{\Lambda}{3},
\ee
where the `dynamical curvature', $\kd$, is again a function of scale, and we have included in
$\kd$ here contributions from both the spatial curvature in the
metric, $\kg$, and the terms in Eq. (\ref{MFE}) that involve
the correlation tensor, $Z^{\alpha \phantom{\beta [ \gamma}
    \mu}_{\phantom{\alpha} \beta[ \gamma \phantom{\mu} \underline{\nu}
      \sigma ]}$.  Only if the contribution from $Z^{\alpha \phantom{\beta [ \gamma}
    \mu}_{\phantom{\alpha} \beta[ \gamma \phantom{\mu} \underline{\nu}
      \sigma ]}$ vanishes do we recover the usual result $\kg=\kd$,
and even in this case we can still have a scale dependence.  More
generally, in spacetimes that are inhomogeneous on small scales, we do
not expect these two `spatial curvature' terms to be equal.
Defining $\Omega_{\kg} = - {\kg}/{a_0^2 H_0^2}$ and $\Omega_{\kd}
= - {\kd}/{a_0^2 H_0^2}$ the macroscopic Friedmann equation then
becomes $1= \Omega_m + \Omega_{\kd} + \Omega_{\Lambda}$,
where $\Omega_m$  and $\Omega_{\Lambda}$ are the usual
expressions for the fraction of the energy content of the Universe in
matter  and the cosmological constant, respectively. Note that
$\Omega_\Lambda$ is a function of smoothing scale whereas $\Lambda$ is
not.  The quantity $\Omega_{\kg}$ does not have to satisfy a constraint of this
kind, as it is now decoupled from the Friedmann equation. Thus we
arrive at a parametrized phenomenological model within which we can
analyze data in order to study the potential observational effects of
averaging. 

\subsection{Observables in the Macroscopic Universe}

We will now consider distance-redshift relations in the FLRW solutions of macroscopic
gravity.  These provide the basis for many key observational tests of
the cosmological background.

First we need to know the
trajectories of photons in the macroscopic geometry.  We will take these to
be null trajectories with respect to the macroscopic metric that has
been constructed to
approximate the distance between two points in spacetime separated by
scales above that of $\Sigma$.  We consider this to be a reasonable
assumption for the average of a large number of photon trajectories,
but note that it will not be true for each individual null curve of
the microscopic spacetime.  If
this assumption is wrong then it could lead to profound differences
with the results of applying Einstein's equations directly to
non-local averaged quantities~\cite{Clarkson:2011br,upcoming}.  Our approach should
therefore be considered a conservative one.

Let us now derive the luminosity distance-redshift relation in the
macroscopic geometry.  Integrating a null trajectory in the geometry
(\ref{MLE}), assuming $\Omega_{\kg}$ and $\Omega_{\kd}$ are constant,
and using the solutions to the macroscopic Friedmann equation
(\ref{FE}), gives
\be
\label{dA}
 d_L (z) =\frac{(1+z)}{H_0  \sqrt{\vert \Omega_{\kg} \vert}}
  f_\kg \left( \int_{\frac{1}{1+z}}^{1}
  \frac{\sqrt{\vert \Omega_{\kg} \vert} da}{\sqrt{\Omega_{\kd} a^2
      +\Omega_{\Lambda} a^4 + \Omega_m a}}\right),
\ee
where $f_\kg(x)=\sinh (x)$, $x$ or $\sin (x)$ when $\kg <0$, $\kg=0$ or
$\kg>0$, respectively.  This expression reduces to the usual one when
$\Omega_{\kg}=\Omega_{\kd}$.  

\section{Data analysis}

We can now compare our averaged models to various cosmological
probes.   

\paragraph*{Hubble Rate.} A vital tool for constraining dark energy is
the local Hubble rate.  We use Hubble rate data from HST
measurements~\cite{Riess:2009pu}, which give $H_0 = 74.2 \pm
3.6$\,km\,s$^{-1}$Mpc$^{-1}$, and we assume Gaussian errors. 

\paragraph*{Cosmic Microwave Background.}

The CMB is well known to tightly constrain spatial
curvature in the standard cosmology.  When $\Omega_{\kg}=\Omega_{\kd}$
constraints from the WMAP 7 year data release \cite{Komatsu:2010fb},
combined with HST \cite{Riess:2009pu} and Baryon Acoustic Oscillation
(BAO) data~\cite{Percival:2009xn}, gives $-0.0133 <\Omega_k < 0.0084$
(95\% CL).  This constraint arises principally from 
the tight bounds on the area distance to the surface of
last scattering, which must be $(1+z_*)d_A(z_*)=14150\pm150\,$Mpc.
We fit three parameters which are predicted by the model: The decoupling epoch ($z_*$), the acoustic scale ($l_A$) and the shift parameter ($R$), which are sufficient to capture the constraints from the CMB~\cite{Komatsu:2010fb}.
To obtain likelihoods, we also use the inverse covariance matrix for the WMAP distance priors for these parameters is given in Table 10 from~\cite{Komatsu:2010fb}.

\paragraph*{Supernovae.}

Another key probe of the large-scale expansion of the
Universe is the observation of type Ia supernovae (SNIa).  These events are
considered to be ``standardizable candles'', in that their absolute
magnitude can be approximated when `stretch' and `color' parameters
have been extracted from fits to light-curve templates.  They
then allow the expansion history of the Universe to be mapped, and are
widely considered to be one of the most compelling sources of evidence
for the existence of dark energy.  We therefore consider them here in
the context of the FLRW solutions to macroscopic gravity, in order to determine
the consequences of allowing $\Omega_{\kg} \neq \Omega_{\kd}$. The
supernova data used in obtaining these constraints are the Union2
data set~\cite{union2}, and the SDSS data set~\cite{SDSSsne}. 

\paragraph*{Baryon Acoustic Oscillations.} Observations of BAOs
provide a direct measurement of the Hubble rate at non-zero
redshifts, and are therefore a powerful tool for constraining dark
energy.  However, the interpretation of BAOs relies on assumptions
about the evolution of structure in the Universe that may not be valid
if averaging is important.  We therefore choose to use BAO data only
sparingly. 
We use the fraction of the comoving sound horizon to volume distance for the two points, at redshifts $z = 0.2$ and $0.35$. The inverse covariance matrix is given by equation~(5)  from~\cite{Percival:2009xn}. 

\subsection{Parameter Constraints}

We use the Monte-Carlo Markov Chain (MCMC) method to obtain
the marginalized errors on the model
parameters from the likelihood function, using the publicity available package COSMOMC~\cite{Lewis02}. 

\begin{figure}[t]
\begin{center}
\includegraphics[height=0.7\columnwidth]{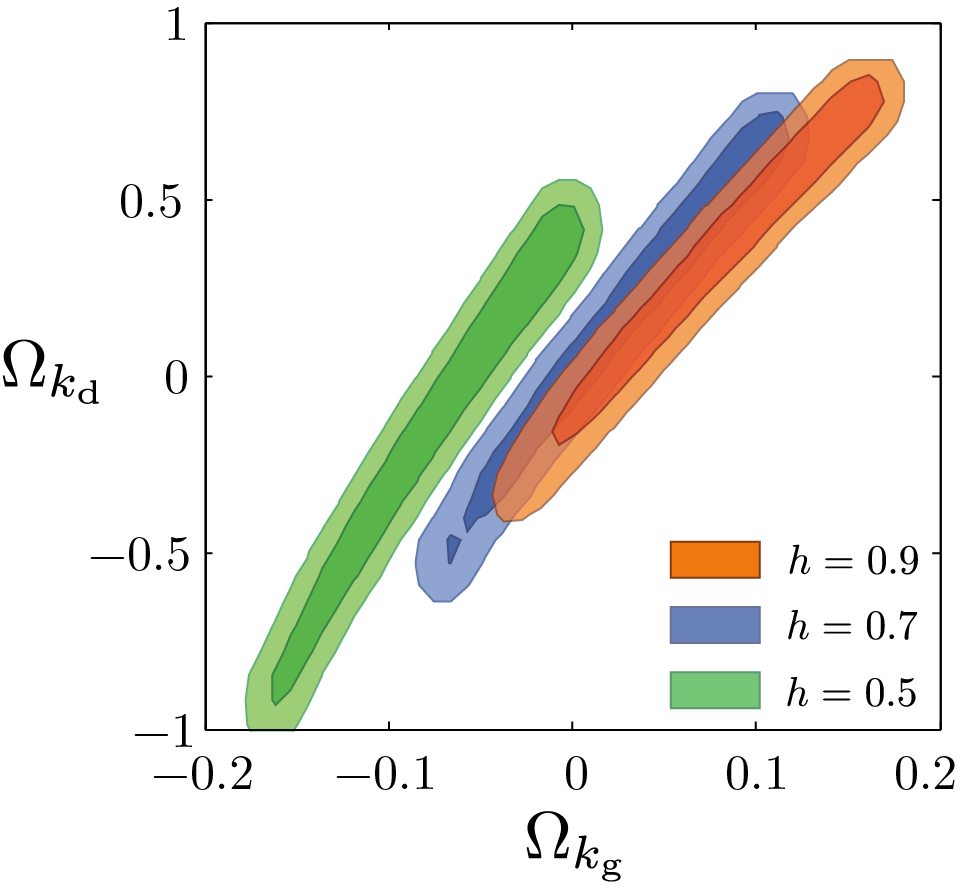}
\caption{Constraints on $\Omega_{\kg}$ and $\Omega_{\kd}$ from the CMB
  for $H_0=50$, $70$ and $90$ km s$^{-1}$ Mpc$^{-1}$, with the other parameters marginalised over. Shaded areas
  are 68\% and 95\% confidence regions.}
\label{hstcmb}
\end{center}
\vspace{-20pt}
\end{figure}

First consider CMB+$H_0$ constraints, as these lead to extremely tight
constraints on spatial curvature in the standard model.  It can be
seen from Fig.~\ref{hstcmb} that when $\Omega_{k_g}\neq \Omega_{k_d}$ the CMB + $H_0$ no longer
constrains spatial curvature significantly, due principally to a
degeneracy between the effects of $\Omega_{\kg}$ and $\Omega_{\kd}$ in
Eq. (\ref{dA}).  In fact, even with $\Lambda=0$ there exist values of
$\kg$ and $\kd$ that satisfy the observations.  These results
significantly weaken a key part of the evidence for both a spatially
flat universe and a non-zero value of $\Lambda$.

\begin{figure*}[ht!]
\begin{center}
\includegraphics[width=1.0\textwidth]{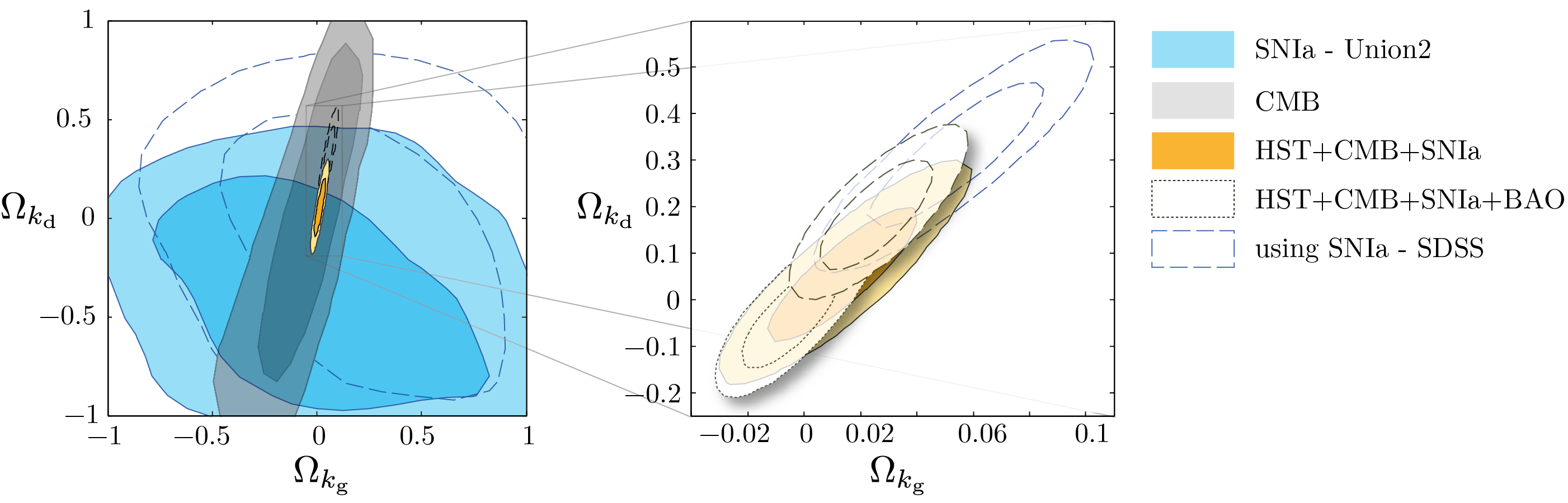}
\caption{Constraints on the two curvature parameters from different
  data sets. In the left-hand panel we show the constraints from the
  CMB (gray) and SNIa (blue for Union2, and hollow and dashed for
  SDSS) separately, as well as the combined constraints including
  HST. In the right-hand panel we show the combined constraints, with
  the smaller lightly shaded regions in the foreground now including
  the BAO. The constraints from the CMB and SNIa individually are
  extremely weak, but tighten when combined with HST data. Note that
  only the geometrical curvature is tightly constrained, and not the
  dynamical one. The SDSS SNIa data can also be seen to inconsistent
  with $\Omega_{k_g}=0$ at greater than 95\% confidence.}
\label{curvatures}
\end{center}
\end{figure*}

\begin{figure}[hb!]
\begin{center}
\includegraphics[width=1.0\columnwidth]{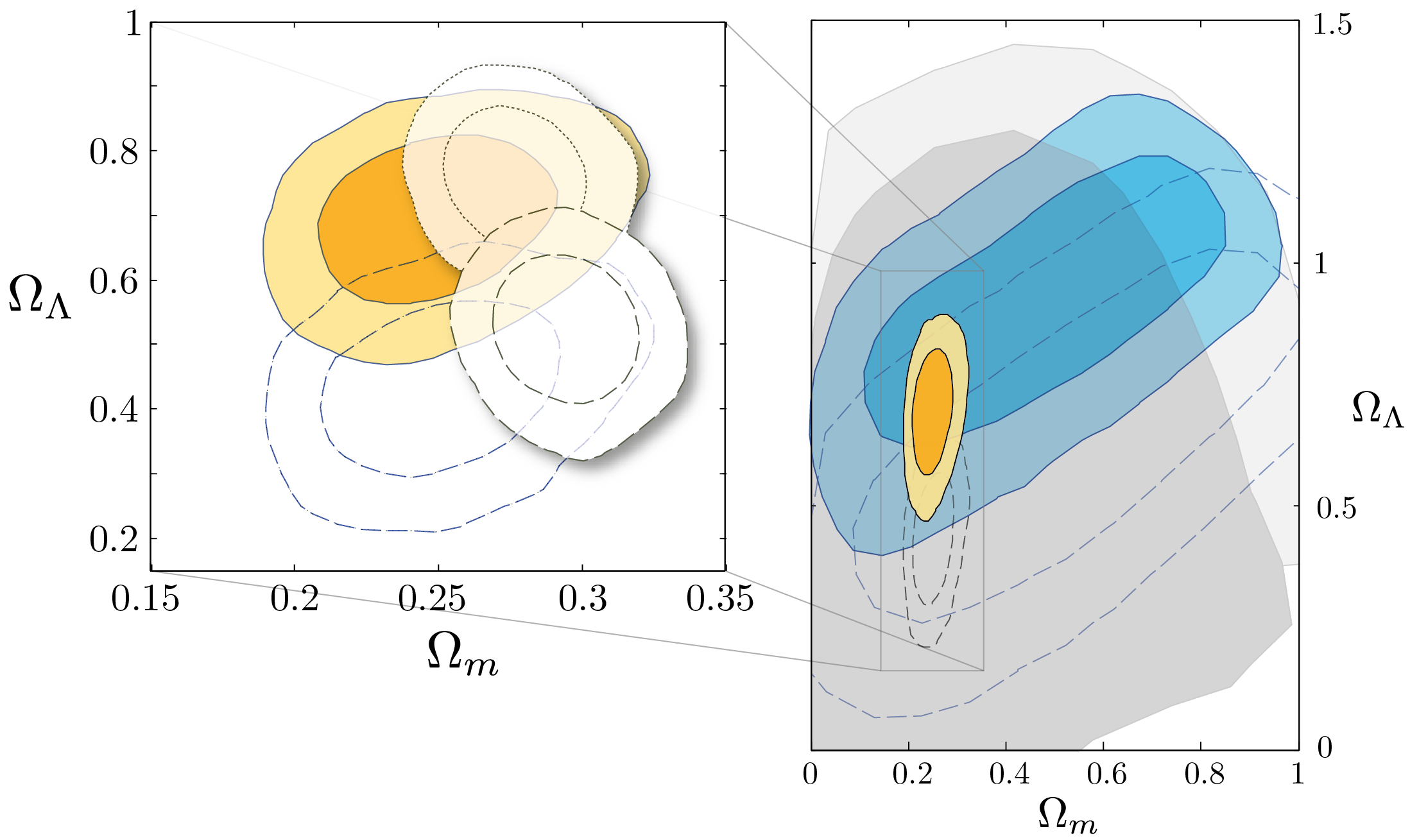}
\caption{Constraints on $\Omega_m$ and $\Omega_\Lambda$ for the same
  data set combinations as in Fig.~\ref{curvatures}. The constraints
  from the CMB and SNIa separately are again much weaker here than in
  the standard case. In particular, $\Lambda=0$ is consistent with the
  CMB data for any $\Omega_m$. Union2 SNIa data, however, still
  clearly gives $\Lambda\neq0$ (for discussion of bias see
  footnote~\cite{footnote1}).}
\label{om-ol}
\end{center}
\end{figure}

In Fig. \ref{curvatures} we show the combined constraints
that can be imposed on $\Omega_{\kg}$ and $\Omega_{\kd}$ using
data from the HST, the WMAP 7 year data of the temperature-temperature
correlations in the CMB, the Union 2 and SDSS SNIa data sets, and the 
constraints on the `volume distance' from the BAOs.  Fig. \ref{om-ol} shows the
constraints available on $\Omega_{m}$ and $\Omega_{\Lambda}$ from the
same data sets.

It is clear that the effect of allowing $\Omega_{\kg}$ and
$\Omega_{\kd}$ to be independent has considerable consequences for
these probability distributions. The marginalized posterior values of
each parameter in the various cases are given in the table.  These
values are considerably different to the case where
$\Omega_{\kg}=\Omega_{\kd}$, which are shown in the same table for
comparison.  It can be seen that the additional freedom 
gained by allowing $\Omega_{\kg} \neq \Omega_{\kd}$ is considerable,
with constraints on $\Omega_{\Lambda}$ and the two $\Omega_{k}$'s being
significantly weaker than in the standard approach.  The combination
of all of these observables, however, still appears to provide strong
evidence for the existence of dark energy, and is still consistent
with a spatially flat universe.  Nevertheless, it is striking that
constraints on $\Omega_{\kg}$ are an order of magnitude tighter than
those on $\Omega_{\kd}$.

\begin{table*}[ht!]
Constraints on curvature and $\Lambda$ with decoupled curvature
parameters, and in the standard model.
\begin{tabular}{c|ccc|cc}
\hline\hline
 Data Sets  &  $  \Omega_{\kg}  $ & $  \Omega_{\kd}  $ & $
 \Omega_{\Lambda}   $ & $  \Omega_{\kd}=\Omega_{\kg}  $ & $
 \Omega_{\Lambda}$    \\[0.5mm] 
 \hline
CMB             & $ -0.053^{+0.152}_{-0.153}$ & $ -0.036^{+0.562}_{-0.572}$ & $ +0.525^{+0.417}_{-0.524}  $
                            & $ -0.069^{+0.109}_{-0.112} $ & $ +0.548^{+0.331}_{-0.303}$ \\
CMB+HST      & $ +0.036^{+0.062}_{-0.064}$ & $ +0.185^{+0.396}_{-0.415}$ & $ +0.564^{+0.415}_{-0.401} $
                             & $ +0.006^{+0.007}_{-0.007}$ & $ +0.746^{+0.023}_{-0.023} $\\
SNIa (Union2)      & $ +0.012^{+0.513}_{-0.485}$ & $ -0.369^{+0.398}_{-0.410}$ & $ +0.902^{+0.189}_{-0.187}  $
                            & $ -0.205^{+0.285}_{-0.282}$ & $ +0.858^{+0.192}_{-0.194} $\\
SNIa (SDSS)          & $ +0.233^{+0.466}_{-0.451}$ & $ -0.173^{+0.492}_{-0.507}$ & $ +0.641^{+0.230}_{-0.225} $
                             & $ +0.073^{+0.301}_{-0.298}$ & $ +0.547^{+0.203}_{-0.204}$\\
CMB+HST+SNIa(Union2)  & $ +0.014^{+0.017}_{-0.017}$ & $ +0.055^{+0.092}_{-0.092}$ & $ +0.695^{+0.080}_{-0.082}$ 
                         & $ +0.005^{+0.007}_{-0.007}$ & $ +0.739^{+0.020}_{-0.021} $\\
CMB+HST+SNIa(SDSS)  & $ +0.054^{+0.020}_{-0.020}$ & $ +0.311^{+0.100}_{-0.101}$ & $ +0.436^{+0.087}_{-0.089} $
                          & $ -0.004^{+0.009}_{-0.009}$ & $ +0.685^{+0.024}_{-0.024} $\\
CMB+HST+SNIa(Union2)+BAO & $ -0.004^{+0.011}_{-0.011}$ & $ -0.033^{+0.070}_{-0.069}$ & $ +0.755^{+0.068}_{-0.070}$ 
                           & $ +0.000^{+0.006}_{-0.006}$ & $ +0.723^{+0.016}_{-0.016} $\\
                           CMB+HST+SNIa(SDSS)+BAO & $
                           +0.026^{+0.012}_{-0.012}$ & $
                           +0.183^{+0.072}_{-0.070}$ & $
                           +0.522^{+0.070}_{-0.073}$ 
                           & $ +0.001^{+0.007}_{-0.006}$ & $ +0.698^{+0.017}_{-0.017} $\\
\hline\hline
\end{tabular}
\end{table*}

\begin{figure}[hb!]
\includegraphics[width=0.7\columnwidth]{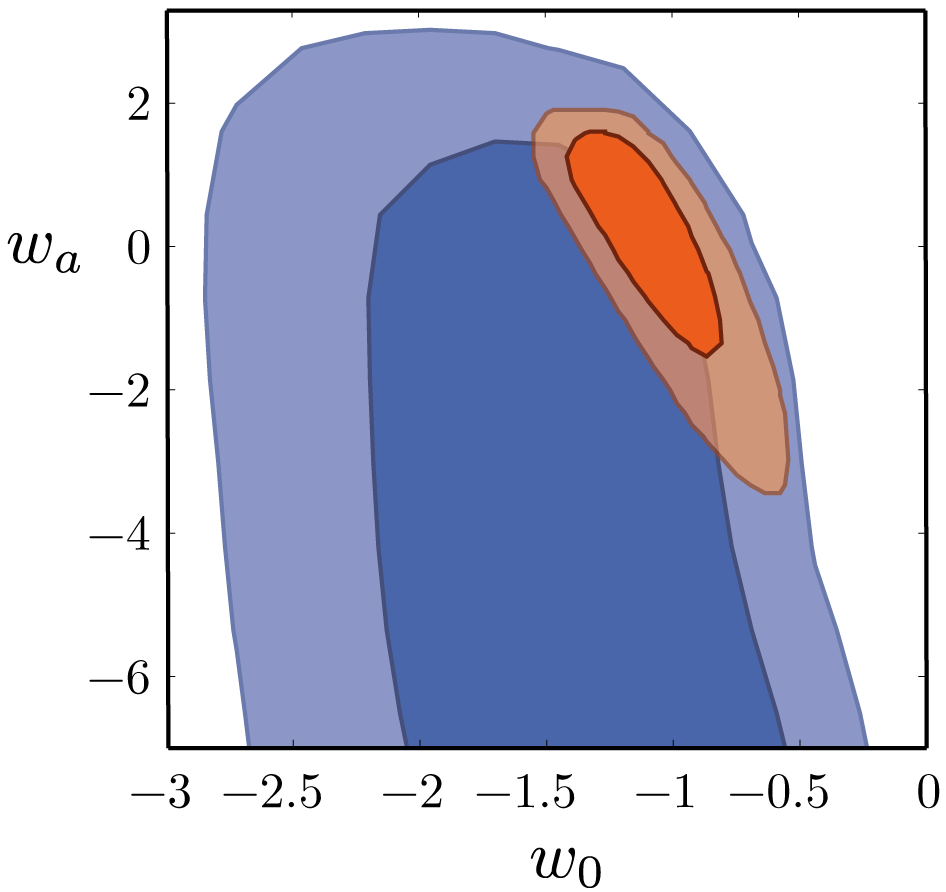}
\caption{Marginalized constraints on the dark energy parameters, with
  (red) and without (blue) assuming $\kg=\kd$. Allowing for effects
  due to averaging removes our ability to constrain the possible
  evolution of the dark energy equation of state.}
\label{w0wa}
\end{figure}

\subsection{Constraints on Dark Energy Dynamics}

Decoupling the geometrical and dynamical spatial curvature parameters
does not appear to allow enough freedom to entirely account for the
dark energy component (which is now signaled almost entirely by
supernova observations, rather than by both CMB+$H_0$ and SNIa observations,
as in the standard model).  Nonetheless, considering averaged
spacetimes does considerably weaken our ability to constrain
dark energy in a meaningful way.  For examples, if we parameterize the
dark energy equation of state as $w(z)=w_0+w_a(1-a)=w_0+w_az/(1+z)$
then CMB+HST+BAO+SNIa(Union2) give only the weak constraint
$w_0=-1.612^{+0.511}_{-0.524}$, with almost no meaningful constraints
on $w_a$ at all.  This can be compared with
$w_0=-1.066^{+0.202}_{-0.197}$ and $w_a=-0.130^{+1.141}_{-1.135}$ when
$\kg=\kd$, as in Fig.~\ref{w0wa}. The uncertainty arising from the
averaging problem is clearly hugely amplified when we try and
constrain the dynamical properties of dark energy.

\subsection{Scale Dependent Curvature Parameters}

As well as $\kg$ and $\kd$ being allowed to be different, a second
consequence of using an averaged geometry is that these parameters
could become scale dependent.  This possibility introduces a 
considerable amount of extra freedom, and hence we consider it
separately here from the more restricted case we have considered so
far (where $\kg$ and $\kd$ are different, but scale independent).

The cosmological probes we have discussed all make
observations over different scales.  The scale of BAOs at the
present time is about $150$ Mpc.  High redshift SNIa (out
to $z\sim 1$) cover a range of scales out to several
Gpc, while the CMB involves making observations on the scale of the
horizon ($\sim14$\,Gpc).  By introducing a scale dependence into $\kg$ and $\kd$
we can therefore potentially re-introduce the effects of spatial
curvature on the scales of SNe and BAOs, while still satisfying the
stringent constraints available on the largest scales from the CMB.
The question then becomes whether or not it is possible for
observations to effectively constrain spatial curvature on different
scales in the Universe.  A positive detection of conflicting
measurements of spatial curvature on different length scales would be
a sure sign of non-trivial averaging effects \cite{COLEYAV,Larena:2008be}.

As we have shown, the CMB may no longer imply spatial flatness or a
non-zero $\Omega_{\Lambda}$ at all (see Fig.~\ref{hstcmb}).  If
spatial curvature on the scale of SNIa observations is allowed to be
independent from the constraints imposed by other observables,
then the evidence available for the existence of a non-zero
$\Omega_{\Lambda}$ is again severely weakened (see Fig.~\ref{om-ol}).
Also, if $\Omega_{\kg}$ and $\Omega_{\kd}$ vary
sufficiently rapidly on scales of interest for SNIa observations then
we can no longer treat them as being simple constants when analyzing the
data. Instead, since we average these observables over the sky, what we
are doing is effectively averaging the geometry out to some redshift.
This should be expected to result in  redshift dependent effective
curvature parameters, and in this case one could end up with curvature
parameters that are effectively functions of radial distance, $k=k(r)$.  

This picture is somewhat similar to LTB void models of the Universe,
where we are at the center of a large spherically symmetric
inhomogeneity.  It is well known that such models are
able to explain the supernova data without evoking the existence of dark
energy \cite{celerier}, but at the expense of strongly violating the Copernican
Principle. In the present interpretation no such violation need occur,
as \emph{all} observers would experience a universe
with $k=k(r)$, with themselves at the center of symmetry.  This would relieve
the key philosophical problem associated with these 
models as an explanation of the data.  It would also relieve the strong
constraints on these models that are available from the kinematic
Sunyaev-Zeldovich effect \cite{kSZ,kSZ2,kSZ3}, as every cluster would
experience (approximately) isotropic CMB radiation, just as we
do. However, such a departure from the usual interpretation of LTB models would
also undoubtedly require revisiting the problem, as the field equations
would be modified by additional terms due to averaging and cannot just be considered as the normal EFE as has been the case so far. In particular, the relation between averaging observables on the sky and averaging the field equations spatially is non-trivial, and extending our ansatz given by Eqs.~(\ref{MLE}) and~(\ref{FE}) to the spherically symmetric case may not be obvious.

\section{Conclusions}

We have presented and constrained models of the large-scale Universe
that result from averaging the geometry of spacetime.  These models have decoupled
spatial curvature parameters in the macroscopic line-element ($\Omega_{\kg}$) and
the Friedmann equation ($\Omega_{\kd}$), and provide a qualitative 
alternative to the standard model of cosmology. They can therefore be used to analyze the
statistical significance of the standard model in a larger space of
models that allows for some of the non-trivial consequences of averaging.

Using HST, CMB, BAO and SNIa data it is clear that the effect of allowing $\Omega_{\kg}$
and $\Omega_{\kd}$ to be independent has considerable consequences for
parameter estimation.  Analysis of the available data shows that the size of the
$68\%$ and $95\%$ confidence regions of $\Omega_{\Lambda}$,
$\Omega_{\kg}$ and $\Omega_{\kd}$ are all much larger than in the
standard model.  There are even tantalizing hints that the data may
favor $\Omega_{\kg} \neq \Omega_{\kd}$ (the combination of HST, CMB
and SDSS SNIa data excludes $\Omega_{\kg} = \Omega_{\kd}$  at the
$95\%$ confidence level).  However, while the evidence for $\Omega_{\Lambda} \neq 0$ available from
individual observables can be considerably reduced, the combination
of SN, CMB, HST and BAO data still provides strong evidence for the
existence of dark energy, as long as $\Omega_{\kg}$ and $\Omega_{\kd}$
are scale independent universal constants.  Relaxing this last
assumption makes combining observables on different scales a much more
complicated problem, and it is highly probable that such additional
freedom will significantly weaken the evidence for $\Omega_{\Lambda}
\neq 0$.  

The way that this situation should be modelled, and constrained
with data, is still an open problem. Relating observables to spatial averages is known to be non-trivial, and is sometimes described as `dressing' the cosmological parameters~\cite{BC1,BC2,Wi1}. Although our ansatz given by Eqs.~(\ref{MLE}) and~(\ref{FE}) is well motivated by macroscopic gravity, other averaging schemes can be used to motivate other phenomenological models, such as in~\cite{Larena:2008be} where Buchert's scheme was used to motivate a time dependent curvature parameter. Other ways of relating average quantities to observables also exist~\cite{LNW,Wi2,WB}, and different observational constraints arise depending on the method used. It remains an open problem to decide on the `correct' way to go about this.

Finally, we have shown that introducing uncertainty due to averaging
into our models of the Universe dramatically weakens the constraints
that can be imposed on the equation of state of dark energy, and we expect this result to be robust to the averaging scheme used.  As a
consequence, it is therefore necessary to understand and
incorporate the effects of averaging in general relativity into our
models if we are to begin attempting precision cosmology.

\section*{Acknowledgements} TC acknowledges the support of the STFC and
the BIPAC, AC acknowledges the support of the NSERC, and RC and CC
acknowledge funding from the NRF (South Africa). Numerical analysis
was performed using facilities at the Centre for High Performance
Computing, South Africa.


\begin{thebibliography}{abc}
\bibitem{footnote1} One should bear in mind here that the Union2 data
  makes use of the SALT light--curve
  fitter~\cite{salt}, which requires ``stretch'' and ``color''
  parameters to be extracted from the model that is being fitted, as
  well as the ``intrinsic error''.  For
the case of the Union2 data these parameters are taken from fitting to
the $\Lambda$CDM model, such that $\chi_{\rm reduced}^2=1$.  This
clearly biases fits against alternative models, such as the one being
considered here, although the consequences of this are
generally thought to be small as long as the shape of the Hubble
diagram is similar to that of $\Lambda$CDM (as is the case here).  For
an example of the effect this can have on models that are not similar
to $\Lambda$CDM see, e.g., \cite{cfl,SW}.
\end{thebibliography}
\end{document}